\documentclass[dvinames]{josr}

\usepackage{microtype}
\usepackage{xcolor}
\usepackage{graphicx}
\usepackage{tikz}
\usetikzlibrary{positioning,arrows,shapes}
\tikzset{
	every picture/.style={/utils/exec={\sffamily}},
	every matrix/.style={ampersand replacement=\&, rounded corners=10pt},
	every node/.style = {font=\small, inner sep = 3},
	>=latex
}
\usepackage{natbib}

\usepackage{enumitem}
\setlist{nosep}

\newcommand{\datadep}[1]{\texttt{datadep"{}#1"{}}}

\usepackage{cleveref}

\begin{document}

\title{DataDeps.jl: Repeatable Data Setup \\ for Replicable Data Science}
\author{Lyndon White, %
	Roberto Togneri, %
	Wei Liu, %
	Mohammed Bennamoun%
	\\ 
	\url{lyndon.white@research.uwa.edu.au}, \\%
	\url{{roberto.togneri,wei.liu,mohammed.bennamoun}@uwa.edu.au}\\
	The University of Western Australia,\\
	35 Stirling Highway, Crawley, Western Australia
}

\maketitle

\begin{abstract}
	We present DataDeps.jl: a julia package for the reproducible handling of static datasets to enhance the repeatability of scripts used in the data and computational sciences.
	It is used to automate the data setup part of running software which accompanies a paper to replicate a result.
	This step is commonly done manually, which expends time and allows for confusion.
	This functionality is also useful for other packages which require data to function (e.g. a trained machine learning based model).
	DataDeps.jl simplifies extending research software by automatically managing the dependencies and makes it easier to run another author's code, thus enhancing the reproducibility of data science research.
\end{abstract}

\textbf{keywords}
data management;  reproducible science; continuous integration, software practices, dependency management, open source software, \mbox{JuliaLang}

\section{Introduction}

In the movement for reproducible sciences there have been two key requests upon authors:
\textbf{1.} Make your research code public, \textbf{2.} Make your data public \citep{lookafterdata}.
In practice this alone is not enough to ensure that results can be replicated.
To get another author's code running on a your own computing environment is often non-trivial.
One aspect of this is data setup: how to acquire the data, and how to connect it to the code.

DataDeps.jl simplifies the data setup step for software written in Julia \citep{Julia}.
DataDeps.jl follows the unix philosophy of doing one job well.
It allows the code to depend on data, and have that data automatically downloaded as required.
It increases replicability of any scientific code that uses static data (e.g. benchmark datasets).
It provides simple methods to orchestrate the data setup: making it easy to create software that works on a new system without any user effort.
While it has been argued that the direct replicability of executing the author's code is a poor substitute for independent reproduction \citep{drummond2009replicability},
we maintain that being able to run the original code is important for checking, for understanding, for extension, and for future comparisons.

\citet{VabdewakkeReproduceableResearch} distinguishes six degrees of replicability for scientific code.
The two highest levels
require that ``The results can be easily reproduced by an independent researcher with at most 15 min of user effort''.
One can  expend much of that time just on setting up the data.
This involves reading the instructions, locating the download link, transferring it to the right location, extracting an archive, and identifying how to inform the script as to where the data is located.
These tasks are automatable and therefore should be automated, as per the practice  ``Let the computer do the work'' \citep{10.1371/journal.pbio.1001745}.

DataDeps.jl handles the data dependencies, while Pkg\footnote{\url{https://github.com/JuliaLang/Pkg.jl}} 
and BinDeps.jl,\footnote{\url{https://github.com/JuliaLang/BinDeps.jl}}
 (etc.) handle the software dependencies.
This makes automated testing possible, e.g., using services such as 
TravisCI\footnote{\url{https://travis-ci.org/}} or AppVeyor.\footnote{\url{https://ci.appveyor.com/}}
Automated testing is already ubiquitous amongst julia users, but rarely for parts where data is involved.
A particular advantage over manual data setup, is that automation allow scheduled tests for URL decay \citep{wren2008url}.
If the full deployment process can be automated, given resources, research can be fully and automatically replicated on a clean continuous integration environment.

\subsection{Three common issues about research data}
\label{sec:issues}
DataDeps.jl is designed around solving common issues researchers have with their file-based data.
The three key problems that it is particularly intended to address are:
\begin{description}
	\item[Storage location:] Where do I put it? \label{itm:where}
	Should it be on the local disk (small) or the network file-store (slow)?
	If I move it, am I going to have to reconfigure things?
	\item[Redistribution:] I don't own this data, am I allowed to redistribute it? \label{itm:ownredistribute} 
	How will I give credit, and ensure the users know who the original creator was?
	\item[Replication:] How can I be sure that someone running my code has the same data?
	What if they download the wrong data, or extract it incorrectly?
	What if it gets corrupted or has been modified and I am unaware?
\end{description}

\section{DataDeps.jl}

\subsection{Ecosystem}

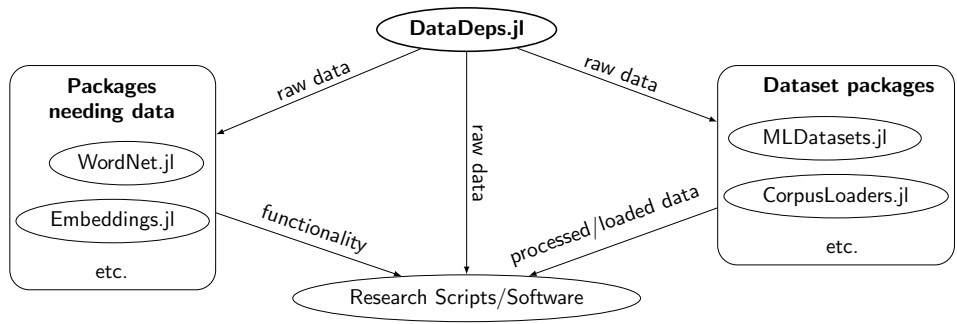
\begin{figure}
	\centering
	\resizebox{!}{0.15\paperheight}{%
		\begin{tikzpicture}[]
		\node[ellipse, thick, draw, inner sep = 3] (datadeps) {\textbf{DataDeps.jl}};
		\node[matrix, xshift=3cm, below right = 0.5 of datadeps, draw, column sep=15, row sep=7] (datasetpackages) {
			\node{\textbf{Dataset packages}}; \\
			\node[draw,ellipse]{MLDatasets.jl};\\
			\node[draw,ellipse]{CorpusLoaders.jl}; \\
			\node{\phantom{MMM} etc. \phantom{M}};\\
		};
		\node[matrix, xshift=-3cm, below left = 0.5 of datadeps, draw, column sep=15, row sep=7] (packages) {
			\node{\textbf{\shortstack{Packages\\needing data}}};\\
			\node[draw,ellipse]{WordNet.jl}; \\
			\node[draw,ellipse]{Embeddings.jl}; \\
			\node{\phantom{MM} etc. \phantom{MM}};\\
		};
		\node[ellipse, draw, inner sep = 3, below = 4 of datadeps] (researchcode) {Research Scripts/Software};
		
		\path[->,draw] (datadeps) edge node[sloped,above]{raw data} (packages);
		\path[->,draw] (packages) edge node[sloped,above]{functionality} (researchcode);
		
		\path[->,draw] (datadeps) edge[above] node[sloped,above]{raw data} (datasetpackages);
		\path[->,draw] (datasetpackages) edge[above] node[sloped,above, xshift=-0.2cm, yshift=0.1cm]{processed/loaded data} (researchcode);
		
		\path[->,draw] (datadeps) edge node[sloped,above]{raw data} (researchcode);	
		\end{tikzpicture}
	}
	\caption{The current package ecosystem depending on DataDeps.jl. 
		%The referenced packages currently exist and are using DataDeps.jl.
		\label{fig:eco}}
\end{figure}

DataDeps.jl is part of a package ecosystem as shown in \Cref{fig:eco}.
It can be used directly by research software, to access the data they depend upon for e.g. evaluations.
Packages such as MLDatasets.jl\footnote{\url{https://github.com/JuliaML/MLDatasets.jl}}
 provide more convenient accesses with suitable preprocessing for commonly used datasets.
These packages currently use DataDeps.jl as a back-end.
Research code also might use DataDeps.jl indirectly by making use of packages, such as WordNet.jl\footnote{\url{https://github.com/JuliaText/WordNet.jl}}
  which currently uses DataDeps.jl to ensure it has the data it depends on to function (see \Cref{sec:research-tool-wordnetjl}); 
or Embeddings.jl which uses it to load pretrained machine-learning models.
Packages and research code alike depend on data, and DataDeps.jl exists to fill that need.

\subsection{Functionality}
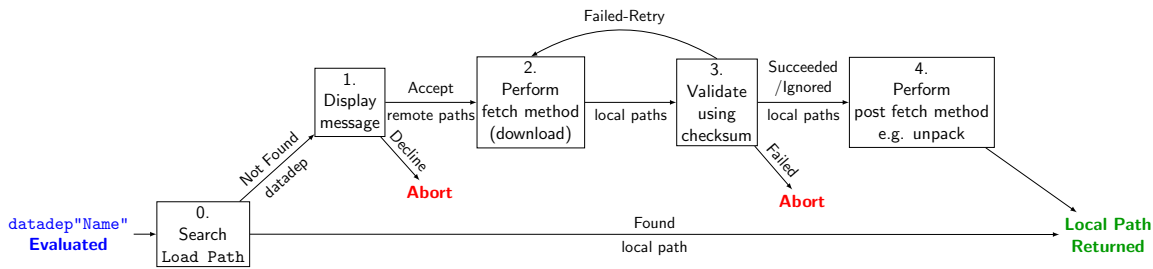
\begin{figure}
	\resizebox{\textwidth}{!}{
		\begin{tikzpicture}[every text node part/.style={align=center}, auto,node distance=2, ->,
		every edge/.append style={every node/.style={font=\footnotesize}}]

		\node[blue](start) {\datadep{Name}\\ \textbf{Evaluated}}; 
		
		\node[draw, rectangle, right= 0.5 of start] (search) {0.\\Search\\ \texttt{Load Path}};
		\node[draw, rectangle, above right= of search] (msg) {1.\\Display\\ message};
		\node[draw, rectangle, right= of msg] (fetch) {2.\\Perform\\ fetch method\\ (download)};
		\node[draw, rectangle, right= of fetch] (checksum) {3.\\Validate\\ using\\ checksum};
		\node[draw, rectangle, right= of checksum] (postfetch) {4.\\Perform \\post fetch method\\ e.g. unpack};
		\node[green!60!black, below right=of postfetch](end) {\textbf{Local Path}\\ \textbf{Returned}};
		
		\path (start) edge (search);
		\path (search)  edge node[sloped,below]{datadep} node[sloped,above]{Not Found}  (msg);
		\path (msg)	edge[below] node{remote paths} node[above]{Accept}  (fetch);
		\path (fetch)	edge[below] node{local paths}  (checksum);
		\path (checksum)	edge node[below]{local paths} node[above]{Succeeded\\/Ignored} (postfetch);
		\path (postfetch)	edge (end);
		\path (search) edge node[above]{Found} node[below]{local path} (end);

		\path (checksum.north) edge[bend right] node[above] {Failed-Retry} (fetch.north);
		
		\node[red, below right = 0.5 of msg, yshift=-0.6cm] (err1) {\textbf{Abort}};
		\path (msg) edge node[above, sloped]{Decline} (err1);
		
		\node[red, below right = 0.5 of checksum, yshift=-0.6cm] (err2) {\textbf{Abort}};
		\path (checksum) edge node[above, sloped]{Failed}  (err2);
		\end{tikzpicture}
	}
	\caption{The process that is executed when a data dependency is accessed by name.		
		 \label{fig:block}
	}
\end{figure}

Once the dependency is declared, data can accessed by name using a datadep string written \datadep{Name}.
This can treated just like a filepath string, however it is actually a string macro.
At compile time it is replaced with a block of code which performs the operation shown in \Cref{fig:block}.
This operation always returns an absolute path string to the data, even that means the data must be download and placed at that path first.

\noindent DataDeps.jl solves the issues in \Cref{sec:issues} as follows:
\begin{description}
	\item[Storage location:] A data dependency is referred to by name, which is resolved to a path on disk by searching a number of locations. The locations search is configurable.
	\item[Redistribution:] DataDeps.jl downloads the package from its original source so it is not redistributed. A prompt is shown to the user before download, which can be set to display information such as the orignal author and any papers to cite etc.
	\item[Replication:] when a dependency is declared, the creator specified the URL to fetch from and post fetch processing to be done (e.g. extraction). This removed the chance for human error.
	To ensure the data is exactly as it was originally checksum is used.
\end{description}

\noindent DataDeps.jl is primarily focused on public, static data.
For researchers who are using private data, or collecting that data while developing the scripts, a manual option is provided; which only includes the \textbf{Storage Location} functionality. They can still refer to it using the \datadep{Name}, but it will not be automatically downloaded.
During publication the researcher can upload their data to an archival repository and update the registration.

\subsection{Similar Tools}
Package managers and build tools can be used to create adhoc solutions, but these solution will often be harder to use and fail to address one or more of the concerns in \Cref{sec:issues}.
Data warehousing tools, and live data APIs; work well with continuous streams of data;
but they are not suitable for simple static datasets that available as a collection of files.

Quilt\footnote{\url{https://github.com/quiltdata/quilt}} is a more similar tool.
In contrast to DataDeps.jl, Quilt uses one centralised data-store, to which users upload the data, and they can then download and use the data as a software package.
%Uploading the data to the centralised Quilt data store raises furth \textbf{Redistribution} issues, further it is a single point of failure, that can not be easily replaced.
It does not directly attempt to handle any \textbf{Storage Location}, or \textbf{Redistribution} issues.
%If the Quilt server goes down all packages depending on quilt break, and it can not be fixed without recreating all of Quilt's  infrastructure
%If an original URL referenced using DataDeps.jl goes down, only software using that URL breaks, and it can be fixed by rehosting the data at a new URL and updating.
%
Quilt does offer some advantages over DataDeps.jl:
excellent convenience methods for some (currently only tabular) file formats, 
and also handling data versioning.
At present DataDeps.jl does not handle versioning, being focused on static data.

\subsection{Quality Control}

Using AppVeyor and Travis CI testing is automatically performed using the latest stable release of Julia, for the Linux, Windows, and Mac environments.
The DataDeps.jl tests include unit tests of key components, as well as comprehensive system/integration tests of different configurations of data dependencies.
These latter tests also form high quality examples to supplement the documentation for users to looking to see how to use the package.
The user can trigger these tests to ensure everything is working on their local machine by the standard julia mechanism: running \texttt{Pkg.test(``DataDeps'')} respectively.

The primary mechanism for user feedback is via Github issues on the repository.
Bugs and feature requests, even purely by the author, are tracked using the Github issues.

\section{Availability}
\subsection{Operating system}
DataDeps.jl is verified to work on Windows 7+, Linux, Mac OSX.

\subsection{Programming language}
Julia v0.6, and v0.7 (1.0 support forthcoming).

\subsection{Dependencies}
DataDeps.jl's dependencies are managed by the julia package manager.
It depends on SHA.jl for the default generation and checking of checksums; 
on Reexport.jl to reexport SHA.jl's methods;
and on HTTP.jl for determining filenames based on the HTTP header information.

\section*{List of contributors}
\begin{itemize}
	\item Lyndon White (The University of Western Australia) Primary Author
	\item Christof Stocker (Unaffiliated), Contributor, significant design discussions.
	\item Sebastin Santy (Birla Institute of Technology and Science), Google Summer of Code Student working on DataDepsGenerators.jl
	
\end{itemize}
%\textcolor{blue}{Please list anyone who helped to create the software (who may also not be an author of this paper), including their roles and affiliations.}

\subsection{Software location:}

\begin{description}[noitemsep,topsep=0pt]
	\item[Name:] oxinabox/DataDeps.jl
	\item[Persistent identifier:] \url{https://github.com/oxinabox/DtaDeps.jl/}
	\item[Licence:] MIT
	\item[Date published:] 28/11/2017
	\item[Documentation Language] English
	\item[Programming Language] Julia
	\item[Code repository] GitHub
\end{description}

\section{Reuse potential}
%\textcolor{blue}{Please describe in as much detail as possible the ways in which the software could be reused by other researchers both within and outside of your field. This should include the use cases for the software, and also details of how the software might be modified or extended (including how contributors should contact you) if appropriate. Also you must include details of what support mechanisms are in place for this software (even if there is no support).}

DataDeps.jl exists only to be reused, it is a ``backend'' library.
The cases in which is should be reused are well discussed above.
It is of benefit to any application, research tool, or scientific script that has a dependency on data for it's functioning or for generation of its result.

DataDeps.jl is extendible via the normal julia methods of subtyping, and composition.
Additional kinds of \texttt{AbstractDataDep} can be created, for example to add an additional validation step, while still reusing the behaviour defined.
Such new types can be created in their own packages, or contributed to the open source DataDeps.jl package.

Julia is a relatively new language with a rapidly growing ecosystem of packages.
It is seeing a lot of up take in many fields of computation sciences, data science and other technical computing.
By establishing tools like DataDeps.jl now, which support the easy reuse of code,
we hope to promote greater resolvability of packages being created later.
Thus in turn leading to more reproducible data and computational science in the future.

\subsection{Case Studies}\label{sec:case-studies}
\paragraph{Research Paper: \citet{White2015BOWgen}}
We criticize our own prior work here, so as to avoid casting aspersions on others. We consider it's limitations and how it would have been improved had it used DataDeps.jl.
Two version of the script were provided\footnote{Source code and data provided at \url{http://white.ucc.asn.au/publications/White2016BOWgen/}}
one with just the source code, and the other also including 3GB of data.
It's license goes to pains to explain which files it covers and which it does not (the data), and to explain the ownership of the data.
DataDeps.jl would avoid the need to include the data, and would make the ownership clear during setup.
Further sharing the source code alone would have been enough, the data would have been downloaded when  (and only if) it is required.
The scripts themselves have relative paths hard-coded. If the data is moved (e.g. to a larger disk) they will break.
Using DataDeps.jl to refer to the data by name would solve this.

\paragraph{Research Tool: WordNet.jl}\label{sec:research-tool-wordnetjl}
WordNet.jl is the Julia binding for the WordNet tool \citep{miller1995wordnet}.
As of PR \#8\footnote{\url{https://github.com/JuliaText/WordNet.jl/pull/8}} it now uses DataDeps.jl.
It depends on having the WordNet database.
Previously, after installing the software using the package manager,
the user had to manually download and set this up.
%Any packages depending on WordNet.jl also would inherit this manual setup problem.
%Now it is handled automatically.
%
The WordNet.jl author previously had concerns about handling the data.
Including it would inflate the repository size, and result in the data being installed to an unreasonable location.
They were also worried that redistributing would violate the copyright.
The manual instructions for downloading and extracting the data included multiple points of possible confusion.
The gzipped tarball must be correctly extracted.
The user must know to pass in the \emph{grand-parent} directory of the database files.
Using DataDeps.jl all these issues have now been solved.

\section*{Acknowledgements}
Thank particularly to Christof Stocker, the creator of MLDatasets.jl (and numerous other packages), in particular for his bug reports, feature requests and code reviews; and for the initial discussion leading to the creation of this tool.

\section*{Competing interests}
The authors declare that they have no competing interests.

\section {Concluding Remarks} 
DataDeps.jl aims to help solve reproducibility issues in data driven research by automating the data setup step. 
It is hoped that by supporting good practices, with tools like DataDeps.jl, now for the still young Julia programming language 
better scientific code can be written in the future . 

\newpage
%\printbibliography
\bibliographystyle{plainnat}
\bibliography{master}

\begin{thebibliography}{8}
\providecommand{\natexlab}[1]{#1}
\providecommand{\url}[1]{\texttt{#1}}
\expandafter\ifx\csname urlstyle\endcsname\relax
  \providecommand{\doi}[1]{doi: #1}\else
  \providecommand{\doi}{doi: \begingroup \urlstyle{rm}\Url}\fi

\bibitem[Bezanson et~al.(2014)Bezanson, Edelman, Karpinski, and Shah]{Julia}
Jeff Bezanson, Alan Edelman, Stefan Karpinski, and Viral~B. Shah.
\newblock {J}ulia: A fresh approach to numerical computing.
\newblock 2014.
\newblock URL \url{http://arxiv.org/abs/1411.1607}.

\bibitem[Drummond(2009)]{drummond2009replicability}
Chris Drummond.
\newblock {Replicability} is not reproducibility: nor is it good science.
\newblock \emph{Proceedings of the Evaluation Methods for Machine Learning
  Workshop at the 26th ICML}, 2009.
\newblock URL \url{http://www.site.uottawa.ca/ICML09WS/papers/w2.pdf}.

\bibitem[Goodman et~al.(2014)Goodman, Pepe, Blocker, Borgman, Cranmer, Crosas,
  Di~Stefano, Gil, Groth, Hedstrom, Hogg, Kashyap, Mahabal, Siemiginowska, and
  Slavkovic]{lookafterdata}
Alyssa Goodman, Alberto Pepe, Alexander~W. Blocker, Christine~L. Borgman, Kyle
  Cranmer, Merce Crosas, Rosanne Di~Stefano, Yolanda Gil, Paul Groth, Margaret
  Hedstrom, David~W. Hogg, Vinay Kashyap, Ashish Mahabal, Aneta Siemiginowska,
  and Aleksandra Slavkovic.
\newblock Ten simple rules for the care and feeding of scientific data.
\newblock \emph{PLOS Computational Biology}, 10\penalty0 (4):\penalty0 1--5, 04
  2014.
\newblock \doi{10.1371/journal.pcbi.1003542}.

\bibitem[Miller(1995)]{miller1995wordnet}
George~A Miller.
\newblock Wordnet: a lexical database for english.
\newblock \emph{Communications of the ACM}, 38\penalty0 (11):\penalty0 39--41,
  1995.

\bibitem[Vandewalle et~al.(2009)Vandewalle, Kovacevic, and
  Vetterli]{VabdewakkeReproduceableResearch}
P.~Vandewalle, J.~Kovacevic, and M.~Vetterli.
\newblock Reproducible research in signal processing.
\newblock \emph{IEEE Signal Processing Magazine}, 26\penalty0 (3):\penalty0
  37--47, May 2009.
\newblock ISSN 1053-5888.
\newblock \doi{10.1109/MSP.2009.932122}.

\bibitem[White et~al.(2016)White, Togneri, Liu, and Bennamoun]{White2015BOWgen}
Lyndon White, Roberto Togneri, Wei Liu, and Mohammed Bennamoun.
\newblock Generating bags of words from the sums of their word embeddings.
\newblock In \emph{17th International Conference on Intelligent Text Processing
  and Computational Linguistics (CICLing)}, 2016.

\bibitem[Wilson et~al.(2014)Wilson, Aruliah, Brown, Chue~Hong, Davis, Guy,
  Haddock, Huff, Mitchell, Plumbley, Waugh, White, and
  Wilson]{10.1371/journal.pbio.1001745}
Greg Wilson, D.~A. Aruliah, C.~Titus Brown, Neil~P. Chue~Hong, Matt Davis,
  Richard~T. Guy, Steven H.~D. Haddock, Kathryn~D. Huff, Ian~M. Mitchell,
  Mark~D. Plumbley, Ben Waugh, Ethan~P. White, and Paul Wilson.
\newblock Best practices for scientific computing.
\newblock \emph{PLOS Biology}, 12\penalty0 (1):\penalty0 1--7, 01 2014.
\newblock \doi{10.1371/journal.pbio.1001745}.
\newblock URL \url{https://doi.org/10.1371/journal.pbio.1001745}.

\bibitem[Wren(2008)]{wren2008url}
Jonathan~D Wren.
\newblock Url decay in medline: a 4-year follow-up study.
\newblock \emph{Bioinformatics}, 24\penalty0 (11):\penalty0 1381--1385, 2008.
\newblock URL
  \url{https://academic.oup.com/bioinformatics/article/24/11/1381/191025}.

\end{thebibliography}

%
%
%\vspace{2cm}
%
%\rule{\textwidth}{1pt}
%
%{ \bf Copyright Notice} \\
%Authors who publish with this journal agree to the following terms: \\
%
%Authors retain copyright and grant the journal right of first publication with the work simultaneously licensed under a  \href{http://creativecommons.org/licenses/by/3.0/}{Creative Commons Attribution License} that allows others to share the work with an acknowledgement of the work's authorship and initial publication in this journal. \\
%
%Authors are able to enter into separate, additional contractual arrangements for the non-exclusive distribution of the journal's published version of the work (e.g., post it to an institutional repository or publish it in a book), with an acknowledgement of its initial publication in this journal. \\
%
%By submitting this paper you agree to the terms of this Copyright Notice, which will apply to this submission if and when it is published by this journal.

\end{document}